\documentclass[twocolumn,superscriptaddress,showpacs]{revtex4}
\usepackage{epsfig}
\usepackage{amsbsy}
\usepackage{amsmath}
\usepackage{amsfonts}
\usepackage{graphicx}

\begin{document}

\newcommand{\nn}{\nonumber}
\newcommand{\dg}{^\dagger}
\newcommand{\bra}[1]{\langle{#1}|}
\newcommand{\ket}[1]{|{#1}\rangle}
\newcommand{\braket}[2]{\langle{#1}|{#2}\rangle}

\title{Relations for classical communication capacity and entanglement
capability of two-qubit operations}
\author{Dominic W.\ Berry}
\affiliation{Department of Physics and
        Centre for Advanced Computing -- Algorithms and Cryptography,   \\
        Macquarie University,
        Sydney, New South Wales 2109, Australia}
\author{Barry C.\ Sanders}
\affiliation{Department of Physics and
        Centre for Advanced Computing -- Algorithms and Cryptography,   \\
        Macquarie University,
        Sydney, New South Wales 2109, Australia}
\date{\today}

\begin{abstract}
Bipartite operations underpin both classical communication and entanglement
generation. Using a superposition of classical messages, we show that the
capacity of a two-qubit operation for error-free entanglement-assisted
bidirectional classical communication can not exceed twice the entanglement
capability. In addition we show that any bipartite two-qubit operation can
increase the communication that may be performed using an ensemble by twice the
entanglement capability.
\end{abstract}
\maketitle

Classical communication can only occur between two systems via an interaction
-- that is, a bipartite or multipartite operation. Similarly, interactions
are essential for generating entanglement between two systems. We
establish relations between the capacities for bipartite operations to perform
these two tasks.

In the static case there is equality between entanglement and shared classical
information. The Schmidt decomposition of a bipartite state $\ket{\Phi} =
\sum_{n=1}^N \sqrt{\lambda_n}\ket{\varphi_n}\ket{\chi_n}$, with
$\ket{\varphi_n}$ and $\ket{\chi_n}$ orthonormal bases for the two systems $A$
and $B$, has an entropy of entanglement given by $E=-\sum\lambda_n \log_2
\lambda_n$ \cite{Ben96}. This entropy of entanglement is also
the mutual information. That is, if system $A$ is measured in the basis
$\ket{\varphi_n}$ and system $B$ is measured in the basis $\ket{\chi_n}$, then
identical results will be obtained. The two sets of measurement results will
then share mutual information $E$. The mutual information, which is classical,
equals the entropy of entanglement, which is quantum. Of course, this mutual
information cannot be used for communication as the measurement results are
obtained in a completely random way, and not on the basis of pre-existing
information.

It is known that there is equality for some simple operations. The CNOT
operation can increase entanglement by a maximum of one ebit (an ebit is the
entanglement contained in one Einstein-Podolsky-Rosen, or EPR, pair). Similarly
it is possible to use a CNOT to communicate one bit in each direction
simultaneously \cite{collins}. Analogously the SWAP gate may produce a maximum
of two ebits or perform two bits of communication in both directions
simultaneously using superdense coding \cite{eisert,super}. In these two cases,
the classical communication capacity in each direction is identical to the
entanglement capability. It is reasonable to hypothesise that this may be true
for all two-qubit operations.

To establish relations, it is useful to consider the entanglement that is
generated by a superposition of classical messages. We employ this concept to
prove an inequality between the entanglement capability and the error-free
classical communication capacity. This result is compared with the recent result
of Bennett \emph{et al.}\ (BHLS) \cite{bennett} where the classical
communication is not assisted by entanglement; i.e.\ entanglement is not
provided as a ``free resource''. We then derive initial ensembles such that the
communication that can be performed using these ensembles may be increased by
the entanglement capability via a two-qubit operation.

We adopt the BHLS terminology, and their manuscript provides an excellent
background on the mathematical foundations for studying entanglement capability
and classical communication capacity. The maximum increase in entanglement that
may be produced by a single implementation of the transform $U$ is denoted
$E_U$. The case of a single implementation of $U$ is sufficient to establish
this bound as it is not possible to generate more entanglement per operation via
multiple operations \cite{bennett}. Here we specifically address the case that
entanglement is quantified by the entropy of entanglement. We employ base-2
logarithms so $E_U$ is in units of ebits.

In addition, we adopt the BHLS definitions for the asymptotic communication
capacities. The bidirectional communication is described by the pair of numbers
$(R_{\rightarrow},R_{\leftarrow})$. The pair $(R_{\rightarrow},R_{\leftarrow})$
is said to be ``achievable'' if, for any $\epsilon>0$, there exists a $t$ such
that it is possible to communicate $tR_{\rightarrow}$ bits from Alice to Bob and
$tR_{\leftarrow}$ bits from Bob to Alice with fidelity $1-\epsilon$ via $t$
applications of $U$ interspersed with local unitary operations. The explicit
mathematical expression for this is given by BHLS. The total capacity is then
defined by
\begin{equation}
C_+^E = \sup \left\{ R_{\rightarrow} + R_{\leftarrow} : 
(R_{\rightarrow},R_{\leftarrow}) {\rm ~is~achievable~by~} U \right\}
\end{equation}
Here we use the superscript $E$ to indicate that any arbitrary amount of
entanglement is allowed to assist communication. The case where no additional
entanglement is allowed to assist communication is denoted by $C_+$ (we do not
consider this case).

Here we consider the case of error-free communication ($\epsilon=0$) for
simplicity. We denote the capacity for error-free communication by $\bar C_+^E$.
It is shown in Ref.\ \cite{eps} that the same results are obtained for
imperfect communication in the limit $\epsilon\to 0$. In order to derive an
inequality for $\bar C_+^E$, we consider a quantum superposition of classical
messages. This technique is similar to that applied by BHLS to derive the
inequality $E_U \ge C_+$. Let us consider a protocol that transmits $n_a$ bits
from Alice to Bob and $n_b$ bits from Bob to Alice, with $t$ uses of $U$. By
taking sufficiently large $t$, the total bidirectional communication per
operation, $(n_a+n_b)/t$, can be made arbitrarily close to $\bar C_+^E$.

We consider an input state $\ket{x}_{A_1}\ket{y}_{B_1}\ket{\psi}_{A_2B_2}$,
where subsystem $A_1$ contains Alice's message $x$ and subsystem $B_1$ contains
Bob's message $y$. $A_2$ and $B_2$ are auxiliary subsystems possessed by Alice
and Bob, respectively, and may contain as much entanglement as is necessary in
order to perform the entanglement assisted communication. Via $t$ applications
of $U$, together with local unitary transformations, we obtain the output state
$\ket{\eta_{xy}}=\ket{y}_{A_1}\ket{x}_{B_1}\ket{c_{xy}}_{A_2B_2}$. That is, the
message $y$ is transferred to Alice, $x$ is transferred to Bob, and the
auxiliary subsystems $A_2$ and $B_2$ are left in a state which may, in general,
depend on $x$ and $y$.

The change in the entanglement is then
\begin{equation}
\Delta E_{xy} = S\left[ {\rm Tr}_{B_2}(\ket{c_{xy}}_{A_2B_2}) \right] -
S\left[ {\rm Tr}_{B_2}(\ket{\psi}_{A_2B_2} ) \right]
\end{equation}
for $S(\rho)=-{\rm Tr}(\rho \log \rho)$, and adopting the convention that
${\rm Tr}_{X}(\ket{\psi}) \equiv {\rm Tr}_{X}(\ket{\psi}\bra{\psi})$.

Now we may consider the case that the input $A_1$ is maximally entangled with
another ancilla $A_3$, and similarly for $B_1$. Then the input state is
\begin{equation}
2^{-(n_a+n_b)/2}\sum_{xy} \ket{x}_{A_1}\ket{x}_{A_3}\ket{y}_{B_1}\ket{y}_{B_3}
\ket{\psi}_{A_2B_2}
\end{equation}
and the output state is
\begin{equation}
2^{-(n_a+n_b)/2}\sum_{xy} \ket{y}_{A_1}\ket{x}_{A_3}\ket{x}_{B_1}\ket{y}_{B_3}
\ket{c_{xy}}_{A_2B_2}.
\end{equation}
The increase in the entanglement for $t$ applications of $U$ is then
\begin{align}
\Delta E & = S\big[ 2^{-(n_a+n_b)} \sum_{xy} \ket{y}_{A_1} \braket{y}{x}_{A_3}
\bra{x} {\rm Tr}_{B_2} (\ket{c_{xy}}_{A_2B_2}) \big] \nn \\ & ~~~ -S\big[
{\rm Tr}_{B_2}(\ket{\psi}_{A_2B_2}) \big] \nn \\
& = (n_a+n_b) + 2^{-(n_a+n_b)} \sum_{xy} S\big[{\rm Tr}_{B_2}
(\ket{c_{xy}}_{A_2B_2}) \big] \nn \\ &~~~ -S\big[{\rm Tr}_{B_2}
(\ket{\psi}_{A_2B_2}) \big] \nn \\
& = (n_a+n_b) + 2^{-(n_a+n_b)} \sum_{xy} \Delta E_{xy}.
\end{align}
For sufficiently large $t$, we may obtain $(n_a+n_b)/t \approx \bar C_+^E$ with
arbitrary accuracy. In addition let $E_U^-$ denote the maximum amount by which
$U$ can decrease the entanglement. We therefore obtain the inequality
$\Delta E/t \ge \bar C_+^E - E_U^-$, implying
\begin{equation}
E_U+E_U^- \ge \bar C_+^E .
\end{equation}

For the case of two-qubit operations, we may simplify this result. In this case
the maximum increase in the entanglement $E_U$ is equal to the maximum decrease
in the entanglement $E_U^-$, which may be shown in the following way. Any
two-qubit interaction $U$ is equivalent, up to local operations, to an operation
of the form \cite{Kraus}
\begin{equation}
\label{simpleU}
U_d = \exp \left[ -i \left( \alpha_1 \sigma_1 \otimes \sigma_1 + \alpha_2
\sigma_2 \otimes \sigma_2 + \alpha_3 \sigma_3 \otimes \sigma_3 \right) \right],
\end{equation}
where $\sigma_k$, for $k\in\{1,2,3\}$ are the Pauli operators.
As entanglement capability and classical communication capacity are independent
of local operations, we may restrict to operations of this form. It is then
simple to show that $U_d^*=U_d\dg=U_d^{-1}$. As discussed in Ref.\ \cite{Kraus},
for any measure of entanglement $E(\ket{\Psi})=E(\ket{\Psi^*})$. This means
that, if the operation $U_d$ acting on the state $\ket{\Psi}$ generates the
maximum increase in entanglement, then this operation performed on the state
$U_d^*\ket{\Psi^*}$ decreases the entanglement by $E_U$. Therefore, the
operation may decrease the entanglement at least as much as it may increase it.
Similarly it is simple to show the converse, and therefore $E_U=E_U^-$. Thus we
find that, for two-qubit operations, we have the inequality
\begin{equation}
\label{myineq}
2E_U \ge \bar C_+^E.
\end{equation}

This means that the sum of the communication capacities in each direction cannot
be greater than twice the entanglement capability. As mentioned above, in the
case of a CNOT or SWAP operation there is equality. That is, the communication
that may be performed in each direction is equal to $E_U$, for a total of
$2E_U$. It would be convenient if we were able to apply a similar argument to
prove the converse. That is, if we were able to decompose any operation creating
entanglement into a superposition of classical messages, then we would be able
to prove $2E_U \le \bar C_+^E$ and therefore equality. Unfortunately, it does
not appear to be possible to establish such a direct equivalence between
classical communication and entanglement generation in this way. Nevertheless we
will show that, via a two-qubit operation with an entanglement capability of
$E_U$, it is possible to increase the communication that can be performed in
each direction using an ensemble by $E_U$ bits.

We will now describe how to obtain initial ensembles for two-qubit operations
such that the Holevo information may be increased by $E_U$ under operation $U$.
In general, for classical message $i$, Bob receives the reduced density matrix
$\rho_i$. When message $i$ is encoded with probability $p_i$, we have the output
ensemble ${\sf E}=\{p_i,\rho_i \}$. Similarly we will denote the ensemble of
pure states shared by Alice and Bob by
${\cal E}=\{ p_i,\ket{\psi_i}_{AB}\}$. We use the subscript $A$ to indicate the
entire subsystem held by Alice and $B$ to indicate the entire subsystem held by
Bob. The Holevo information $\chi$ for ensemble ${\sf E}$ is given by
\begin{equation}
\label{holevo}
\chi({\sf E}) = S\left( \sum_i p_i \rho_i \right) - \sum_i p_i S(\rho_i).
\end{equation}
This communication may be achieved via coding over multiple states
\cite{Holevo}. 

The entanglement-assisted communication capacity in a single direction is
\cite{bennett}
\begin{equation}
C_{\rightarrow}^E = \sup \big\{ R: (R,0) {\rm ~is~achievable~by~}U \big\}.
\end{equation}
As the two-qubit operation is symmetric between Alice and Bob, all results for
this case also apply to communication in the opposite direction,
$C_{\leftarrow}^E$. BHLS show that $C_{\rightarrow}^E$ is given by
\begin{equation}
\label{commun}
C_{\rightarrow}^E = \sup_{\cal E} \big[ \chi({\rm Tr}_{A} U{\cal E})-
\chi({\rm Tr}_{A}{\cal E}) \big].
\end{equation}
Similarly to BHLS we adopt the notation convention
\begin{align}
U{\cal E} & = \{ p_i, U\ket{\psi_i}_{AB} \}, \\
{\rm Tr}_{A}{\cal E} & = \{ p_i, {\rm Tr}_{A}(\ket{\psi_i}_{AB}) \}.
\end{align}

The reason why the communication rate is of the form of a difference in the
Holevo information is that, for any $\epsilon,\delta>0$, for sufficiently
large $n$ it is possible to construct $n$ copies of the ensemble ${\cal E}$
with fidelity $1-\epsilon$ using communication $n[\chi({\rm Tr}_{A}{\cal E})+
\delta]$ \cite{berry,Shor}. In the limit of small $\epsilon$ and $\delta$, this
means that $\chi({\rm Tr}_{A}{\cal E})$ bits are required for construction of
each ensemble ${\cal E}$. The communication that may be performed after the
operation $U$ is then $\chi({\rm Tr}_{A} U{\cal E})$. As communication
$\chi({\rm Tr}_{A}{\cal E})$ was required to construct the ensemble, the
additional communication per operation is as given by Eq.\ (\ref{commun}).

In order to see how to obtain communication equal to $E_U$, note that the second
term on the right-hand side of Eq.\ (\ref{holevo}) is the average of the
entanglement of the coding states. Therefore, if each of the initial
states $\ket{\psi_i}_{AB}$ are chosen such that the entanglement of these
states is decreased by the maximum $E_U$ by operation $U$, then the second term
in (\ref{holevo}) will be decreased by $E_U$ by the operation. If the first term
is constant, then the total communication will be $E_U$.

In order to obtain such an ensemble, let us start with an initial state
$\ket{\Psi}$ such that the entanglement is decreased by the maximum $E_U$ via
operation $U$. We then wish to find a set of operations
$\{ V_i^{A}V_i^{B} \}$, where $V_i^{A}$ are local operations on Alice's
side, $V_i^{B}$ are local operations on Bob's side, and
$V_i\equiv V_i^{A}V_i^{B}$ commutes with $U$. This means that each state
$V_i \ket{\Psi}$ will have its entanglement decreased by $E_U$ for
operation $U$. We also require $\sum_i p_i {\rm Tr}_{A} (V_i \ket{\phi})$ to
be a multiple of the identity for all input states $\ket{\phi}$. Then
\begin{equation}
\sum_i p_i{\rm Tr}_{A} (U V_i\ket{\Psi}) =
\sum_i p_i{\rm Tr}_{A} (V_i U \ket{\Psi}) \propto \openone .
\end{equation}
This means that the first term on the right side of (\ref{holevo}) will be
unchanged by the operation $U$, and therefore that the total one-way
communication possible is $E_U$.

Let us consider the case that, in addition to the two qubits upon which $U$
acts, Alice and Bob each possess one auxiliary qubit. It has been found
\cite{leifer} that the maximal increase in entanglement may be achieved using
only one auxiliary qubit on each side, and no improvements are obtained using
additional auxiliary qubits. We will label Alice's auxiliary qubit 1, the two
qubits upon which $U$ acts that Alice and Bob possess as 2 and 3, respectively,
and Bob's auxiliary qubit as 4. For simplicity, and without loss of generality,
we will restrict to operations $U$ given in the form (\ref{simpleU}). In this
case the operators $\sigma_i^{(2)} \otimes \sigma_i^{(3)}$, for
$i\in\{0,1,2,3\}$, commute with $U$. Here we take the notation convention that
$\sigma_0$ is the identity and the superscripts indicate the qubits upon
which these operators act. Similarly the operators
$\sigma_i^{(1)} \otimes \sigma_i^{(4)}$ commute with $U$. Let us then consider
the set of 16 operators $\{ \sigma_{i'}^{(1)} \sigma_i^{(2)} \sigma_i^{(3)}
\sigma_{i'}^{(4)} \}$. Given a single state $\ket{\Psi}$ for which $U$ decreases
the entanglement by $E_U$, we obtain a set of 16
states $\{\sigma_{i'}^{(1)} \sigma_i^{(2)} \sigma_i^{(3)} \sigma_{i'}^{(4)}
\ket{\Psi}\}$ for which the entanglement will be decreased by $E_U$. In
addition, it is easily shown that, for any
two-qubit state $\rho$,
\begin{equation}
\label{identity}
\sum_{ii'} \sigma_i^{(3)} \sigma_{i'}^{(4)} \rho \sigma_i^{(3)}
\sigma_{i'}^{(4)} \propto \openone.
\end{equation}
This implies that
\begin{equation}
\sum_{ii'} {\rm Tr}_{12} \big(\sigma_{i'}^{(1)} \sigma_i^{(2)} \sigma_i^{(3)}
\sigma_{i'}^{(4)} \ket{\Psi}\big) \propto \openone.
\end{equation}
Thus we find that, by coding with equal probabilities $p_{ii'}=1/16$ each of the
16 states $\sigma_{i'}^{(1)} \sigma_i^{(2)} \sigma_i^{(3)} \sigma_{i'}^{(4)}
\ket{\Psi}$, it is possible to communicate $E_U$ bits in one direction using
operation $U$.

Next we consider the more complicated case of bidirectional communication. Let
$i$ be the classical message encoded by Alice, and $j$ be the classical message
encoded by Bob. The output ensemble is then ${\cal E} = \{p_i,q_j,
\ket{\psi_{ij}}_{AB} \}$. We will denote the bidirectional communication
that it is possible to perform using ensemble ${\cal E}$ by
$\chi^\leftrightarrow ({\cal E})$. We also define
\begin{equation}
\Delta\chi_U^\leftrightarrow = \sup_{\cal E} [\chi^\leftrightarrow (U{\cal E})
- \chi^\leftrightarrow ({\cal E})].
\end{equation}

We again consider a state $\ket\Psi$ such that $U$ decreases the entanglement by
the maximum $E_U$. Alice encodes via the set of 16 operators
$\{\sigma_{i'}^{(1)}\sigma_i^{(2)}\sigma_i^{(3)}\sigma_{i'}^{(4)}\}$, and Bob
encodes via the 16 operators $\{ \sigma_{j'}^{(1)} \sigma_j^{(2)} \sigma_j^{(3)}
\sigma_{j'}^{(4)} \}$. Note that Alice and Bob's operators commute, so the order
in which these operators are applied is irrelevant. We therefore have a total
ensemble of 256 states
\begin{equation}
{\cal E} = \{ p_{ii'}, q_{jj'}, \sigma_{i'}^{(1)} \sigma_i^{(2)}\sigma_i^{(3)}
\sigma_{i'}^{(4)} \sigma_{j'}^{(1)} \sigma_j^{(2)} \sigma_j^{(3)}
\sigma_{j'}^{(4)} \ket\Psi \},
\end{equation}
where we take $p_{ii'}=q_{jj'}=1/16$. As each of the operators applied by Alice
and Bob commutes with $U$ [for $U$ in the form (\ref{simpleU})], the
entanglement will be decreased by $E_U$ for all of these states. 

For this type of bidirectional ensemble, the bidirectional communication that it
is possible to perform is given by \cite{eps}
\begin{equation}
\chi^\leftrightarrow ({\cal E}) = \chi^{\rightarrow}({\rm Tr}_{A}
{\cal E})+\chi^{\leftarrow}({\rm Tr}_{B}{\cal E}),
\end{equation}
where we have defined
\begin{align}
\label{right}
\chi^{\rightarrow}{\sf E} &= \sum_j q_j \left[ S\left( \sum_i p_i \rho_{ij}
\right) - \sum_i p_i S(\rho_{ij})\right], \\
\label{left}
\chi^{\leftarrow}{\sf E}  &= \sum_i p_i \left[ S\left( \sum_j q_j \rho_{ij}
\right) - \sum_j q_j S(\rho_{ij})\right],
\end{align}
for the ensemble ${\sf E}=\{ p_i,q_j,\rho_{ij} \}$.

Now using the relation (\ref{identity}), we find that
\begin{equation}
\sum_{ii'} {\rm Tr}_{12} \big(\sigma_{i'}^{(1)} \sigma_i^{(2)} \sigma_i^{(3)}
\sigma_{i'}^{(4)} \sigma_{j'}^{(1)} \sigma_j^{(2)} \sigma_j^{(3)}
\sigma_{j'}^{(4)} \ket{\Psi}\big) \propto \openone
\end{equation}
and
\begin{equation}
\sum_{jj'} {\rm Tr}_{34} \big(\sigma_{i'}^{(1)} \sigma_i^{(2)} \sigma_i^{(3)}
\sigma_{i'}^{(4)} \sigma_{j'}^{(1)} \sigma_j^{(2)} \sigma_j^{(3)}
\sigma_{j'}^{(4)} \ket{\Psi}\big) \propto \openone.
\end{equation}
These relations mean that the first terms in the sums in (\ref{right}) and
(\ref{left}) are constant under the action of operation $U$. As the entanglement
$S(\rho_{ij})$ decreases by $E_U$ for each state, we find that the total
communication that it is possible to perform in each direction using the
ensemble is increased by $E_U$, for a total increase of
$2E_U$.

Thus we have shown the sequence of inequalities $\bar C_+^E\le 2E_U\le\Delta
\chi_U^\leftrightarrow$. From Ref.\ \cite{eps}, the same results are obtained
for asymptotically perfect communication as for error-free communication, so
$C_+^E\le 2E_U\le\Delta\chi_U^\leftrightarrow$.
We conjecture that, similarly to the case for communication in a
single direction \cite{berry,Shor}, we may obtain $n$ copies of the ensemble
of states ${\cal E}$ using bidirectional communication of
$n[\chi^{\leftrightarrow}({\cal E})+\delta]$. In the limit of large $n$, this
would imply that the ensemble ${\cal E}$ may be constructed using bidirectional
communication per ensemble $\chi^{\leftrightarrow}({\cal E})$. Given this
conjecture, we may achieve an average bidirectional communication of $C_+^E=
\Delta\chi_U^\leftrightarrow$ per operation $U$, using a scheme equivalent to
that considered by BHLS for the case of communication in a single direction.
Together with the above results, this would imply that $C_+^E$ is equal to
twice the entanglement capability of the operation. That is, we may communicate
$E_U$ bits in each direction per operation.

In summary, we have shown that any two-qubit operation~$U$ can increase the
communication that may be performed using an ensemble by an amount equal to the
entanglement capability of $U$. This applies both to cases of communication in
just one direction and communication in both directions simultaneously.
Furthermore, if the conjecture that the initial pure-state ensemble~${\cal E}$
may be constructed using as much classical communication as can be performed
using this ensemble (in the limit of a large number of samples) is valid, then
the total communication that may be performed in each direction per operation
equals the entanglement capability $E_U$ in the limit of a large number of
operations. Even if this conjecture is not valid (which would be very
interesting in itself), we have established that, for arbitrary $U$, there
exists ${\cal E}$ such that the communication that can be performed using this
ensemble increases by $E_U$ after application of $U$ to ${\cal E}$.

\textbf{Acknowledgments:} This project has been supported by an
Australian Research Council Large Grant.


\begin{thebibliography}{}
\bibitem{Ben96} C. H. Bennett {\it et al.}, \pra \textbf{54}, 3824 (1996).
\bibitem{collins} D. Collins, N. Linden, and S. Popescu, \pra {\bf 64}, 032302
(2001).
\bibitem{eisert} J. Eisert {\it et al.},
\pra {\bf 62}, 052317 (2000).
\bibitem{super} C. H. Bennett and S. J. Wiesner, \prl {\bf 69}, 2881 (1992).
\bibitem{bennett} C. H. Bennett {\it et al.}, quant-ph/0205057 (2002).
\bibitem{eps} D. W. Berry and B. C. Sanders, quant-ph/0207065 (2002).
\bibitem{Kraus} B. Kraus and J. I. Cirac, \pra {\bf 63}, 062309 (2001).
\bibitem{Holevo} A. S. Holevo, IEEE Trans. Inf. Theory {\bf 44}, 269 (1998).
\bibitem{Shor} Ref. \cite{bennett} cites a personal communication by
P. W. Shor and a paper in preparation by A. W. Harrow.
\bibitem{berry} D. W. Berry and B. C. Sanders, quant-ph/0209093 (2002).
\bibitem{leifer} M. S. Leifer, L. Henderson, and N. Linden,
quant-ph/0205055 (2002).
\end{thebibliography}
\end{document}